# Role of Magnetic Coupling in Photoluminescence Kinetics of $Mn^{2+}$-doped ZnS Nanoplatelets

*Liwei Dai,[a] Abderrezak Torche,[a] Christian Strelow,[a] Tobias Kipp,[a] Thanh Huyen Vuong,[b] Jabor Rabeah,[b] Kevin Oldenburg,[c] Gabriel Bester,[a] Alf Mews,[a] Christian Klinke,[c,d,e] and Rostyslav Lesyuk,*[c,f]*

[a]*Institute of Physical Chemistry, University of Hamburg, Martin-Luther-King-Platz 6, 20146 Hamburg, Germany*

[b]*Leibniz Institute for Catalysis, 18059 Rostock, Germany*

[c]*Department "Life, Light & Matter", Center for Interdisciplinary Electron Microscopy (ELMI-MV), University of Rostock, Albert-Einstein-Strasse 25, 18059 Rostock, Germany*

[d] *Institute of Physics, University of Rostock, Albert-Einstein-Straße 23, 18059 Rostock, Germany*

[e]*Department of Chemistry, Swansea University – Singleton Park, Swansea SA2 8PP, UK*

[f] *Pidstryhach Institute for Applied Problems of Mechanics and Mathematics of NAS of Ukraine, Naukowa str. 3b, 79060 Lviv & Department of Photonics, Lviv Polytechnic National University, Bandery str. 12, 79000 Lviv, Ukraine*

ABSTRACT

Mn$^{2+}$-doped semiconductor nanocrystals with tuned location and concentration of Mn$^{2+}$ ions can yield diverse coupling regimes, which can highly influence their optical properties such as emission wavelength and photoluminescence (PL) lifetime. However, investigation on the relationship between the Mn$^{2+}$ concentration and the optical properties is still challenging because of the complex interactions of Mn$^{2+}$ ions and the host and between the Mn$^{2+}$ ions. Here, atomically flat ZnS nanoplatelets (NPLs) with uniform thickness were chosen as matrixes for Mn$^{2+}$ doping. Using time-resolved (TR) PL spectroscopy and density functional theory (DFT) calculations, a connection between coupling and PL kinetics of Mn$^{2+}$ ions was established. Moreover, it was found that the Mn$^{2+}$ ions residing on the surface of a nanostructure produce emissive states and interfere with the change of properties by Mn$^{2+}$–Mn$^{2+}$ coupling. In a configuration with suppressed surface contribution to the optical response we show the underlying physical reasons for double and triple exponential decay by DFT methods. We believe that the presented doping strategy and simulation methodology of the Mn$^{2+}$-doped ZnS (ZnS:Mn) system is a universal platform to study dopant location- and concentration-dependent properties also in other semiconductors.

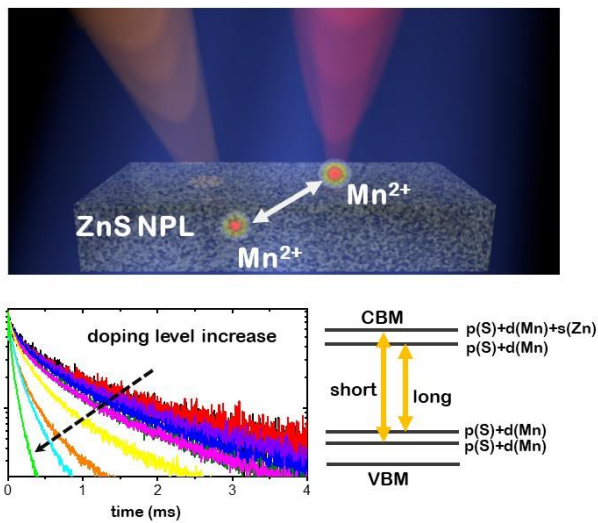

INTRODUCTION

Mn$^{2+}$-doped II−VI semiconductor nanocrystals (NCs) attracted quite some research interest owing to Mn$^{2+}$-related photoluminescence (PL) with distinct features. For instance, the Mn$^{2+}$ emission is mainly unaffected by the host material and therefore noticeably red-shifted with respect to the large band-gap energy of the host materials.[1-2] Consequently, undesired re-absorption losses can be minimized. Additionally, Mn$^{2+}$-doped semiconductor NCs can exhibit relatively high PL quantum yields (QYs). In 2011, Deng et al. reported a simple and "green" phosphine-free method to synthesize high-quality Mn$^{2+}$-doped ZnS quantum rods (QRs) with a PLQY of 45%.[3] Recently, Peng's group developed a new synthetic route for Mn$^{2+}$-doped ZnSe/ZnS core/shell dots with variable Mn$^{2+}$ concentrations.[4] The resulting dots showed a high PLQY of up to 90%. In addition to these superior optical properties, the thermal and environmental stability of semiconductor NCs is found to be enhanced dramatically after Mn$^{2+}$ doping.[5-6] These advantages make Mn$^{2+}$-doped semiconductor NCs attractive for applications in light-emitting diodes (LEDs), biomedical labeling, laser, and other optoelectronic devices based on densely packed NCs.[7-9]

Another characteristic feature of the Mn$^{2+}$ emission is the long decay lifetime (microseconds to milliseconds) due to the spin-forbidden $^4T_1$–$^6A_1$ *d–d* transitions,[10-12] which triggers potential applications in phosphors, efficient extraction of charge carriers, and quantum information processing.[13-14] Many efforts have been made to achieve such a long and mono-exponential decay lifetime.[15-17] The main principle is to disperse Mn$^{2+}$ ions into the host material homogeneously to maintain the tetrahedral coordination (substitutional doping). This, however, requires precise control of the dopant sites. In most of the Mn$^{2+}$-doped systems, the obtained decay lifetime of the Mn$^{2+}$ emission is shortened and multi-exponential due to the formation of magnetic coupling between Mn$^{2+}$ ions or Mn$^{2+}$ ions residing on the surface of a nanoparticle, which is especially observed at high doping concentrations and/or in ultra-small hosts.[18-21] In this case, the decay kinetics are complicated since $Mn^{2+}$ ions with different degrees of coupling may generate new relaxation pathways.[15] Thus, it remains challenging to assign the decay channels to different light-emitting $Mn^{2+}$ species.

For a given $Mn^{2+}$-doped system, the decay lifetime of the $Mn^{2+}$ emission is strongly distance-dependent: isolated $Mn^{2+}$ ions (from each other) usually emit with a lifetime in the range of milliseconds, while the lifetime shortens gradually with decreasing $Mn^{2+}$−$Mn^{2+}$ distance.[15, 22-23] $Mn^{2+}$-doped ZnS nanoplatelets (ZnS:Mn NPLs) offer a new platform to study conveniently the relationship between the light-emitting $Mn^{2+}$ species and their decay lifetime and to develop prototypes for efficient down-converters with controlled PL kinetics. Very thin and atomically flat NPLs are also an ideal model system to address the properties of $Mn^{2+}$ ions occupying surface locations. Here, we attempt to connect the location- and concentration-dependent time-resolved (TR) PL signatures of ZnS:Mn NPLs both by experiments and density functional theory (DFT) calculations. Further, we explain the interplay between different decay channels observed in the experiments. The experimental results reveal that all PL decay curves of $Mn^{2+}$-doped samples can be well fitted by a triexponential function. In combination with the possibility to post-synthetically overgrow the surface by colloidal atomic layer deposition (c-ALD), the surface effect can be addressed and deciphered and – as shown later – substantially suppressed giving the possibility to study the $Mn^{2+}$–$Mn^{2+}$ interaction effect. With the assist of DFT calculations, the origins of the slow, intermediate and fast decays can be assigned to corresponding transitions (*p−d- or s–p-*character) of isolated, weakly coupled, and strongly coupled $Mn^{2+}$ ions.

## RESULTS AND DISCUSSIONS

To study the decay kinetics of light-emitting $Mn^{2+}$ species, we here choose ZnS NPLs in wurtzite (WZ) phase as the host matrix for several seasons. First, zinc and manganese ions have similar ionic radii (74 vs 80 pm),[24-25] and thus $Mn^{2+}$ ions can be easily incorporated into Zn-containing matrixes. Second, the size of the host can significantly influence the optical properties of the dopant, which may result in some size-dependent properties.[26-27] Since quasi-two-dimensional (quasi-2D) NPLs are known to have an atomically flat structure (internal uniformity) and their thickness is monodisperse in one batch of synthesis (external uniformity),[28] they can effectively eliminate the dimensional inhomogeneities. The ZnS:Mn NPLs with different doping levels were synthesized according to a reported method.[29-30] The high-resolution transmission electron microscopy (HRTEM) image (Figure 1a) shows that the resulting platelets with an average lateral size of 15.8 × 6.4 $nm^2$ (Figure S1) display a constant contrast, suggesting an uniform thickness. This is confirmed by the UV−vis absorbance spectrum (Figure 1b, blue) of ZnS:Mn NPLs showing a strong and narrow excitonic peak at 4.38 eV. The large blue-shift (ca. 0.6 eV) of the excitonic peak in comparison to the band-gap of the bulk ZnS (3.6−3.8 eV)[31-32] reveals the existence of a strong quantum confinement effect due to the ultrathin thickness of the NPLs. The steady-state PL spectrum (Figure 1b, red) shows a broad $Mn^{2+}$ emission band centered at 2.08 eV, revealing a red-shift (2.3 eV) in emission with respect to the ZnS absorption. Such a large shift can effectively avoid self-absorption of $Mn^{2+}$ emission, which is beneficial to the following investigation of the $Mn^{2+}$ emission kinetics.

A series of ZnS:Mn NPL samples with increasing $Mn^{2+}$ doping concentration were studied ranging from 0.03 to 16.6% of Mn:Zn atomic ratio (by IPC-OES) which roughly corresponds to an average of 1.4 – 793 dopants per NPL. The average number of $Mn^{2+}$ ions per NPL is referred to our previous calculations.[30] TEM images of three representative samples (low, intermediate, and high doping level, Figure S2) show that the shape of the products remains unchanged as the doping level changes, manifesting that doping of $Mn^{2+}$ ions has no obvious influence on the shape of the host (ZnS). The UV−vis absorbance spectra for the set of the samples (Figure S3, solid line) display almost the same spectral profile with an excitonic peak fixed at 4.38 eV. It is worth noting that the excitonic peak broadens slightly at stronger doping levels (16.6%), which could be ascribed to the influence of $Mn^{2+}$ ions on the electronic structure of the host NPLs.[15] The corresponding PL spectra (Figure S3, dashed line) show that the center of the $Mn^{2+}$ emission band red-shifts ($\Delta E$ = 50 meV) as the doping level increases, reflecting the increased $Mn^{2+}-Mn^{2+}$ coupling. The $Mn^{2+}$ emission kinetics was studied by TRPL measurements with a kHz laser system (see experimental section for details). For all samples, the $Mn^{2+}$ PL decay curves exhibit a multi-exponential decay behavior (Figure 1c), suggesting an inhomogeneous local environment of $Mn^{2+}$ ions in the NPLs. These decay curves can be well fitted by a triexponential function:[33]

$$I(t) = y_0 + A_1\exp(-t/\tau_1) + A_2\exp(-t/\tau_2) + A_3\exp(-t/\tau_3), \quad (1)$$

where $\tau_1$, $\tau_2$, and $\tau_3$ are the time constants and $A_1$, $A_2$, and $A_3$ are the normalized amplitudes of the components. The detailed fitting parameters are listed in Table S1. The average lifetime ($\tau_{ave}$) can be determined by the equation:[33]

$$\tau_{ave} = (A_1\tau_1^2 + A_2\tau_2^2 + A_3\tau_3^2)/(A_1\tau_1 + A_2\tau_2 + A_3\tau_3). \quad (2)$$

Figure 2a shows that the average lifetime decreases as the doping level increases, which can be attributed to the increasing $Mn^{2+}-Mn^{2+}$ coupling. The strong magnetic coupling between $Mn^{2+}$ ions, evidenced by our previous electron paramagnetic resonance (EPR) results,[30] could partially lift the spin-forbidding nature of the $^4T_1$–$^6A_1$ $d$–$d$ transition, resulting in reduced PL decay lifetime.

The development of the time constants of the three components with doping level is shown in Figure 2b. We conditionally classify the samples by the Mn:Zn atomic ratio ≤ 0.38% (average number of $Mn^{2+}$ ions per NPL ≤ 18) as "Group I" and the rest as "Group II". For Group I, the time constants of the three decay channels (slow, intermediate, and fast) are found to be on the scale of 1−1.6ms, hundreds of µs, and tens of µs, respectively (Figure 2b). With an increase of the doping level in the range 0.03–0.38% (average number of $Mn^{2+}$ ions per NPL from 1.4 to 18), the time constants of the intermediate and fast decay channels keep nearly constant. The time constant of the slow decay channel is rather stable in the range 1−1.6 ms even if a slight reduction is observed. In contrast, the time constant of the slow decay channel in Group II decreases dramatically, while the time constants of the intermediate and fast decay channels are only slightly decreasing as the doping level increases (Figure 2b and c). At the maximal doping level (16.6% or ≈793 $Mn^{2+}$ ions per NPL), the time constant of the slow decay channel reaches the same time scale as that of the intermediate decay channel.

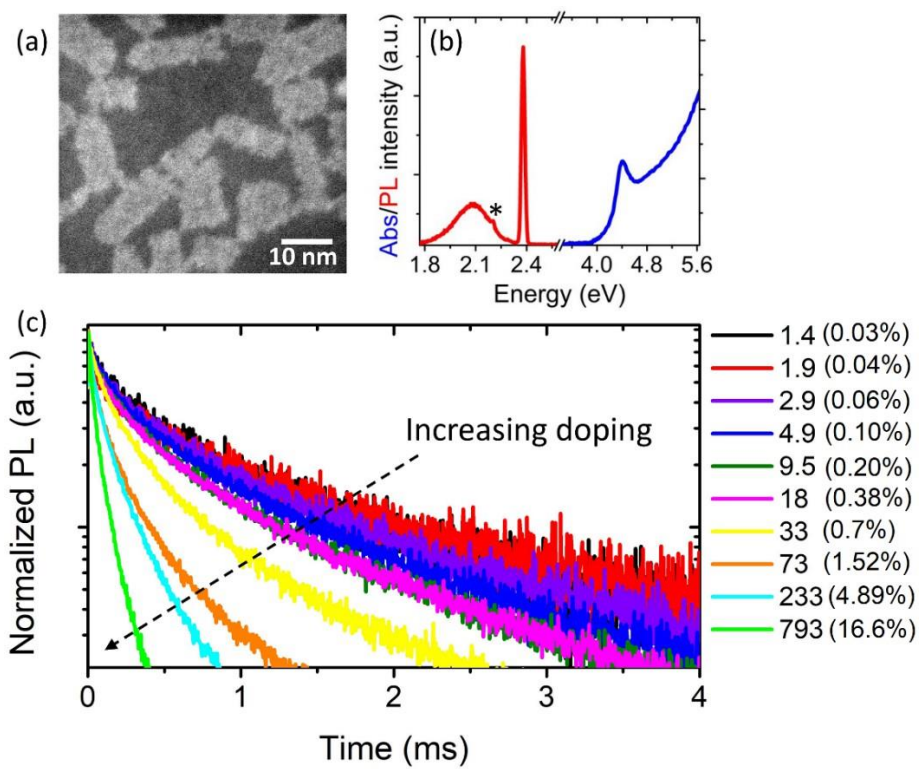

**Figure 1.** (a) HRTEM image of ZnS:Mn NPLs (exemplarily shown for the intermediate doping level of 1.5% or ≈ 73 $Mn^{2+}$ ions per NPL). (b) The corresponding steady-state UV−vis absorbance (blue) and PL (red) spectra. Excitation is at 4.77 eV (260 nm). The scattered light of the lamp {a narrow peak at 2.38 eV (520 nm)} and the solvent (hexane) Raman peaks (marked with an asterisk) are also detected. (c) PL decay curves of ZnS:Mn NPLs with increasing $Mn^{2+}$ doping level detected at the maximum of the $Mn^{2+}$ emission peak with an excitation at 4.35 eV (285 nm).

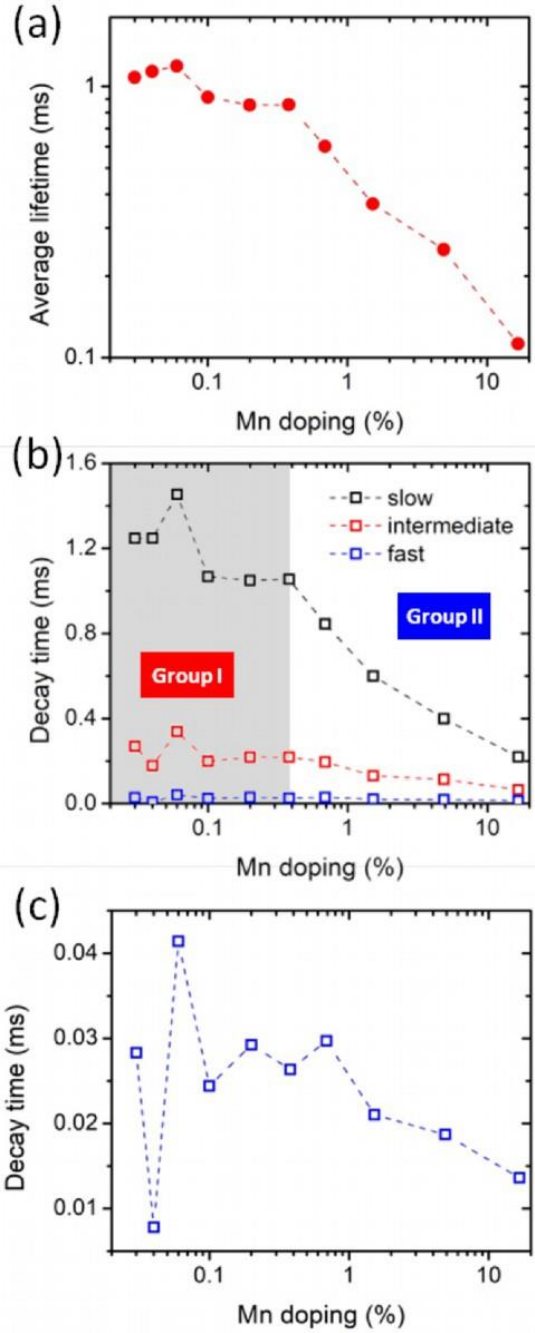

**Figure 2.** Plots of (a) average lifetime and (b) time constants of three decay channels as a function of $Mn^{2+}$ doping level. (c) Magnified view of the time constant of the fast decay channel against the doping level.

In order to investigate the microscopic origin of the three decay channels, DFT calculations were performed. We have used DFT within the local density approximation (LDA) and norm conserving pseudopotential with a cutoff of 30 Ha. A slab containing 288 atoms with a thickness of 1.52 nm was constructed to simulate the ultrathin ZnS:Mn NPLs. Since the $Mn^{2+}$ $d-d$ transition is both orbital- and spin- forbidden, spin polarization and spin-orbit coupling have to be considered in the ground state calculation in order to relax these selection rules and to have a finite lifetime. Figure 3a shows the unit cell used to simulate the $Mn^{2+}$-doped slab together with the resulting DFT-calculated eigen-energies (b) and a schematic of the energy levels in the area around the band gap (c). In Figure 3b, the color bar denotes the $d$-character of the state (higher value means approaching the pure $d$-state). The half-filled and 10-fold degenerate $d$-states of the $Mn^{2+}$ ion split in the ZnS environment because of the spin-orbit, spin polarization and ligand field effects. The filled states reside near the valence-band maximum (VBM) while the empty states reside near the conduction-band minimum (CBM). Note that the splitting between filled and empty $d$-states, which corresponds to the observed orange emission, might be underestimated in our calculation because of the LDA used in DFT that is known to perform less well in the description of localized states. Nevertheless, for lifetime calculations DFT gives reliable results, which are consistent with the experiments as discussed later. In the independent particle approximation (*i.e.* DFT level without excitonic effects) the lifetime of an optical transition can be estimated as[34]:

$$t \text{ (ns)} = \frac{1}{21.42\, n\, E\, |M_{if}|^2}, \qquad (3)$$

where $t$ is the lifetime in nanosecond, $E$ is the $d-d$ transition energy (in Hartree), $n$ is the refractive index (~2.5 for ZnS), and $M_{if}$ is the dipole matrix between in the $i$ and $f$ states in atomic units. The situation where an $Mn^{2+}$ ion substitutes a $Zn^{2+}$ ion in the 288 atom unit cell as depicted in Figure 3a corresponds to a doping of 0.7 % and $Mn^{2+}-Mn^{2+}$ distance of 19.8 Å, for which the $Mn^{2+}$ ions are isolated and can be considered as non-interacting. At this doping level, lifetime calculation (Table 1) shows that the isolated $Mn^{2+}$ ions have two lifetime components, a long one in the range of a few milliseconds (3.2 ms) and a short one in the range of hundreds of microseconds (370 μs). The origin of these two components is schematically depicted in Figure 3c as transitions between filled and empty $d$-states with different hybridization with the host materials. The transition with long component occurs between $p(S) + d(Mn)$ orbits so it is an $p-d$-like transition whereas the short component stems from transition between a $p(S) + d(Mn)$ hybridized orbit and a $p(S) + d(Mn) + s(Zn)$ orbit so it is closer to a $s-p$ transition. It is known that $s-p$ transitions are the strongest, thus, they indeed reflect a fast decay.

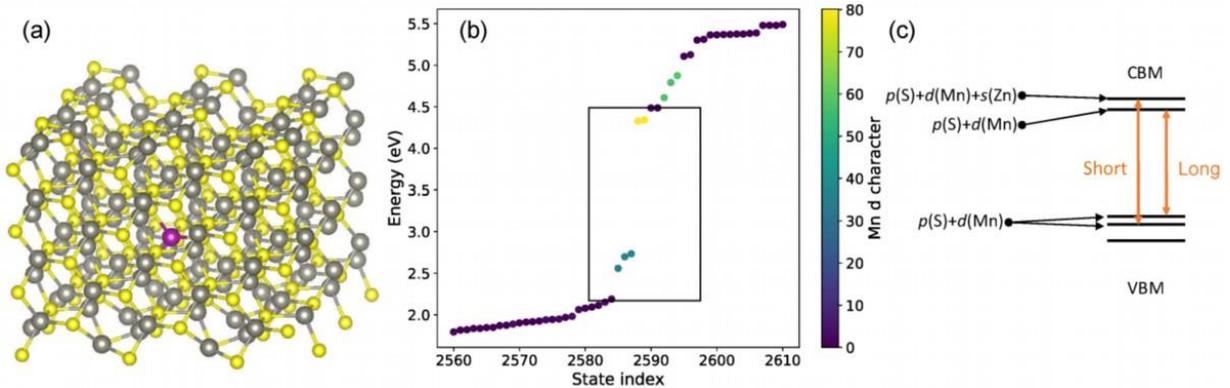

**Figure 3.** (a) The Mn (purple) doped ZnS unit cell (gray and yellow) with 288 atoms used to simulate the platelet. (b) Energy levels of the doped slab in (a) at the DFT-LDA level. The color bar reflects the $d$-character of the $Mn^{2+}$ states. (c) Schematics of the energy levels in (b) around the gap areas showing the origin (orbital hybridization) of the state involved in the $Mn^{2+}$ emission (long and short components) around 2.08 eV.

To trace back the variation of the $Mn^{2+}$ lifetime with doping level within DFT we performed lifetime calculations with 2 $Mn^{2+}$ ions inside the unit cell at different distances. The results are depicted in Table 1. In all cases, two different decay components are revealed. The long component lifetime decreases rather evenly from 3.2 to 1.27 ms with decreasing distance between $Mn^{2+}$ ions. The decrease in the long component lifetime is due to the increased hybridization between $d$-orbitals of the two $Mn^{2+}$ ions, which progressively lifts the selection rule for spin-forbidden transitions and makes the decay of the $Mn^{2+}$ transition more probable and thus faster. Interestingly, the short component first does not change strongly, only for the smallest distance possible in the lattice between the $Mn^{2+}$ ions (3.8 Å), the lifetime drastically shortens by nearly one order of magnitude to 50 μs. The distance of 3.8 Å is close to the dimer formation distance,[35] where the hybridization between the $d$-orbitals is at maximum. We note that the energy splitting between long and short transitions is on the scale of tens of meV (approx. 50 meV for the configuration in Figure 3).

**Table 1.** DFT calculations of $Mn^{2+}$ (long-living and short-living components) PL lifetime as a function of Mn−Mn distance for the isolated, weakly coupled, and strongly coupled $Mn^{2+}$ ions, respectively.

| $Mn^{2+}$ species | Mn–Mn distance (Å) | long (ms) | short (µs) |
|---|---|---|---|
| isolated | 19.8 | 3.20 | 370 |
| weakly coupled | 14.7 | 3.16 | 220 |
| | 12.0 | 2.08 | 270 |
| | 9.5 | 2.01 | 330 |
| | 6.2 | 1.55 | 330 |
| strongly coupled | 3.8 | 1.27 | **50** |

Based on Table 1, $Mn^{2+}$ ions with different $Mn^{2+}-Mn^{2+}$ distances in the ZnS NPLs can be classified into three scenarios: non-interacting isolated $Mn^{2+}$ ions (distance ≥ 19.8 Å), weakly coupled $Mn^{2+}$ ions (3.8 Å < distance < 19,8 Å), and strongly coupled $Mn^{2+}$ ions (distance = 3.8 Å). Although the simulation results reveal that each $Mn^{2+}$ species has only two decay components (long and short), a triexponential function is found to fit the PL decay curves well because ensembles of NPLs were measured. By comparing the results of the experiments and the simulations we find that the time constant of the slow decay channel has a maximum in the doping range of Group I, and then it significantly decreases with increasing doping level. This corresponds to the lifetime variation of the long component of $Mn^{2+}$ ions (from isolated to weakly coupled to strongly coupled $Mn^{2+}$ ions). Thus, we attribute the origin of the slow decay channel to the long-living component of $Mn^{2+}$ ions. Considering that $Mn^{2+}$ ions in Group I are mainly isolated, evidenced by EPR results in Figure S4 and our previous report,[30] the "actual" time constant of the long component of the isolated $Mn^{2+}$ ions can be estimated to be 1.2 ms by averaging the time constants of the slow decay channel of Group I. The time constant of the intermediate decay channel slightly decreases with increasing doping level, basically consistent with the change of the short component as the $Mn^{2+}$ distance decreases from 19.8 to 6.2 Å in simulations. This indicates that the origin of the intermediate decay channel can be ascribed to the short-living component of isolated and weakly coupled $Mn^{2+}$ ions (corresponding to $Mn^{2+}$ distance between 19.8−6.2 Å). Besides, since the time constant of the slow decay channel is close to the time scale of the short component lifetime of the strongly coupled $Mn^{2+}$ ions, it is reasonable to attribute the origin of the fast decay channel to the short-living component of the strongly coupled $Mn^{2+}$ ions. Similarly, the "actual" time constant of the short component of the strongly coupled $Mn^{2+}$ ions is estimated to be 24 μs by averaging the time constants of the fast decay channel of the whole series of the samples (Group I + Group II, Figure 2c). Interestingly, if a third $Mn^{2+}$ ion is added at the nearest distance of 3.8 Å in our model (clustering of Mn), the lifetime of the short-living component decreases further to approx. 22 μs which is even closer to the fitting parameter for the experimental curve.

To study the fraction of the three $Mn^{2+}$ species as a function of doping level, global fitting was performed with sharing the fixed time constants for the isolated $Mn^{2+}$ ions (1.2 ms) and the strongly coupled $Mn^{2+}$ions (24 μs). The time constant for the weakly coupled $Mn^{2+}$ ions was set as a free parameter due to the fact that the average distance between weakly coupled $Mn^{2+}$ ions differs in samples with different doping levels. The fitting parameters are listed in Table S2. In Group I, the fraction of the isolated $Mn^{2+}$ ions is dominant (above 75%) and changes slightly (Figure 4), revealing the relatively stable environment of $Mn^{2+}$ ions. The number of the weakly coupled $Mn^{2+}$ ions increases on the cost of the isolated $Mn^{2+}$ ions whereby the number of the strongly coupled (aggregated) $Mn^{2+}$ ions remains nearly constant. However, apparent changes are observed in Group II. The fraction of the isolated $Mn^{2+}$ ions decreases sharply, which in turn leads to an increasing fraction of the other two $Mn^{2+}$ species.

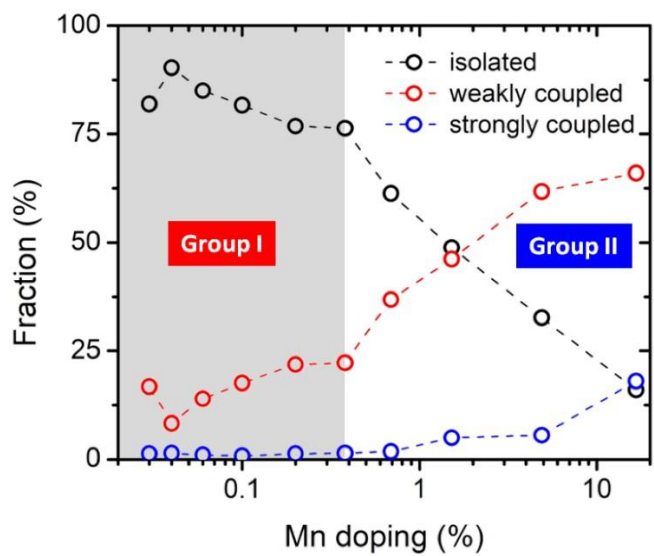

**Figure 4.** Plots of the fraction of three $Mn^{2+}$ species as a function of doping level.

Despite a good qualitative agreement of experimental and theoretical PL lifetime dependencies on the doping level and the distance between $Mn^{2+}$ ions, some quantitative discrepancies are evident. Generally, the experimental decay components tend to be shorter than predicted by the calculations. This points to the presence of additional factors, which might influence the PL decay kinetics. Our previous report[30] shows that the ZnS:Mn NPLs not only have inner $Mn^{2+}$ ions, but also $Mn^{2+}$ ions that reside on the surface of NPLs (surface $Mn^{2+}$ ions). To investigate the influence of the surface $Mn^{2+}$ ions on the PL kinetics, we prepared ZnS:Mn/ZnS core/shell NPLs using the c-ALD method.[36] For that, a typical ZnS:Mn NPL sample with an intermediate doping level (0.7% or ≈33 $Mn^{2+}$ ions per NPL) was used as the core. As schematically shown in Figure 5a, the formed ZnS shell can well passivate the surface $Mn^{2+}$ ions, ensuring that only inner $Mn^{2+}$ ions are present in the platelets after the passivation. The sample prepared with c-ALD exhibits a platelet-like shape (Figure S2, d–f), similar to the core platelets. XRD pattern of the sample (Figure S5) shows that the diffraction peaks correspond to the WZ structure of ZnS (ICPDS 00−080−0007). The shell formation is confirmed by comparing the UV−vis absorbance spectra of samples before and after c-ALD. The absorbance spectrum of ZnS:Mn/ZnS red-shifts compared to that of ZnS:Mn, suggesting an increase in thickness due to the shell formation (Figure 5b). The thickness is anticipated to increase from 1.8 nm [29] to maximally ≈3 nm according to the c-ALD procedure (three cycles of $S^{2-}$ and $Zn^{2+}$ deposition, see *Experimental Section*). It is anticipated, that one c-ALD cycle can produce half a monolayer in the thickness direction. Before studying the PL kinetics of ZnS:Mn/ZnS core/shell NPLs, we tested whether the surface $Mn^{2+}$ ions contribute to the PL, since opposite effects, both PL quenching and clear PL of surface Mn, were reported in different works.[37-40] Figure 5c shows a comparison of steady-state PL spectra between ZnS:Mn core NPLs and ZnS:Mn/ZnS core/shell NPLs. We note that the PL spectrum narrows and blue-shifts when the surface $Mn^{2+}$ ions are covered with a shell. This discloses that in our system, the surface $Mn^{2+}$ ions in NPLs without a shell indeed emit and their emission resides at longer wavelengths than that of the inner $Mn^{2+}$ ions. This observation agrees with results of Sarma's group for CdS NCs[37]. To get more insights to the role of surface dopants, we investigated the PL kinetics by performing TRPL measurements at different emission wavelengths for the core and core/shell samples. Note that these measurements were done with a MHz laser system and an optical chopper (see experimental section for details). The 2D PL decay spectra become more symmetrical and narrower when the surface $Mn^{2+}$ ions are passivated by a ZnS shell (Figure 5d and g), further indicating that the surface $Mn^{2+}$ ions tend to emit at longer wavelengths. For comparison, PL decay curves at three representative emission wavelengths (560, 600, and 640 nm) were recorded (inset in Figure 5e and h). For the core sample, the PL decay curves differ from one wavelength to another (Figure 5e), while the PL decay curves of the core/shell samples remain unchanged (Figure 5h). This reveals the heterogeneous character of the $Mn^{2+}$ emission of the core NPLs due to the presence of the surface $Mn^{2+}$ ions. Besides, the PL decay curve of the core sample shortens with increasing emission wavelength, reflecting that the surface $Mn^{2+}$ ions has a shorter decay lifetime relative to the inner $Mn^{2+}$ ions. We also compared the transient PL spectra recorded at different decay time for the core and core/shell samples. Figure 5f shows a small blue-shift in the PL spectra of the core sample, which is absent in the core/shell sample (Figure 5i). We attribute the blue-shift to the short decay lifetime of the surface $Mn^{2+}$ ions. This is reasonable because the contribution of the surface $Mn^{2+}$ ions to the emission is found to decrease with the decay time (Figure S6). Here we note that the spectral separation of the emission of inner and surface $Mn^{2+}$ ions is about 140 meV noticeably exceeding the calculated spectral separation of the long and short components coming from $Mn^{2+}$–$Mn^{2+}$ coupling (Figure 3b and c). Due to the reduction of the long-wavelength emission intensity originating from the surface $Mn^{2+}$ ions, the integral emission gradually blue-shifts. To clarify the contribution of the surface from the theory side, we also simulated the $Mn^{2+}$ ion residing on the ZnS NPL surface. This results in a decrease of the lifetime approx. by a factor of 3 in comparison to the $Mn^{2+}$ ion that occupies an atomic location inside the NPL. The energy of the transition is further shifted to the red part of the spectrum in qualitative agreement with the experimental results. Note that surface computation was conducted without passivation due to the large number of atoms required in such simulations. We however observed no surface states that are extended over the whole surface, which indicates that passivation most probably plays a minor role.

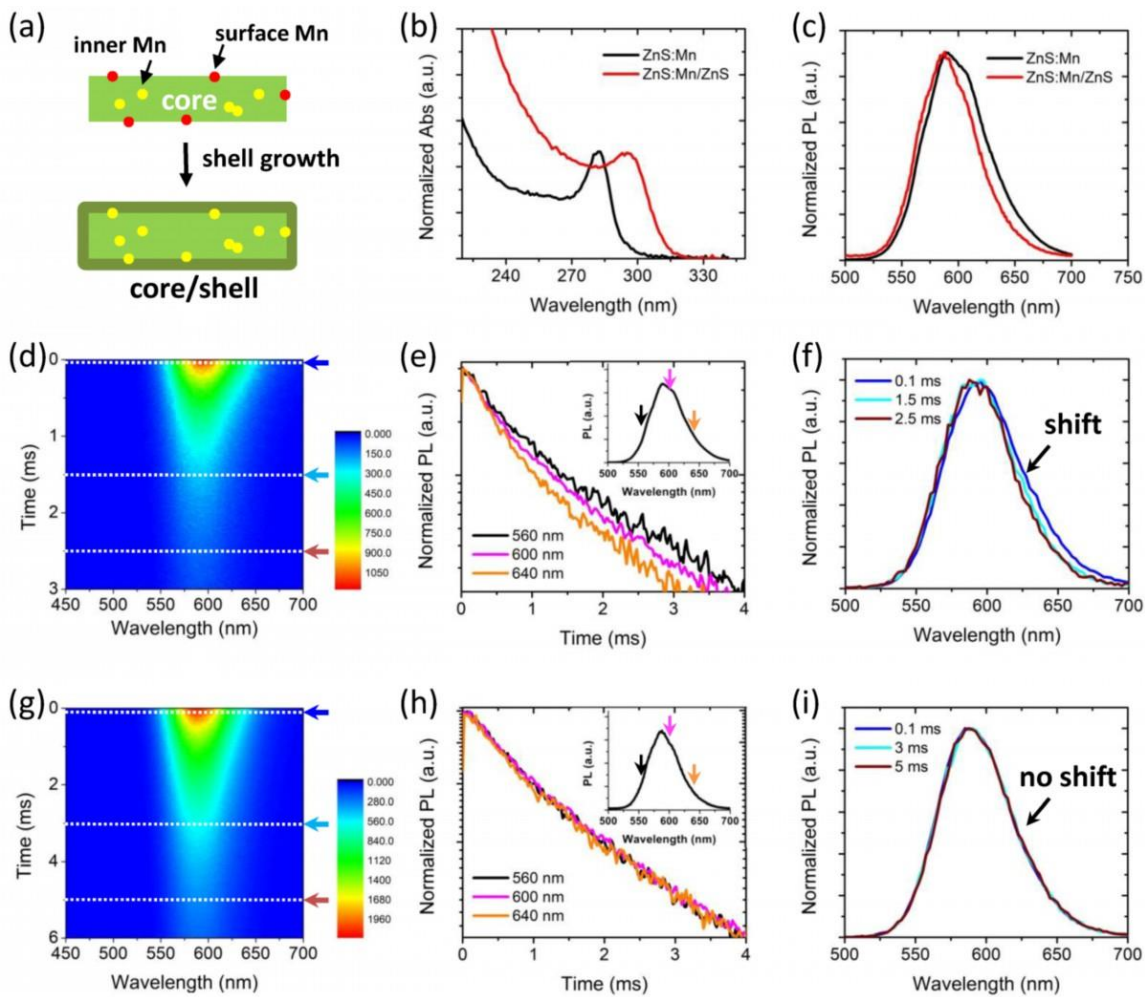

**Figure 5.** (a) Schematic illustration of ZnS shell formation on the ZnS:Mn core NPLs. (b) Steady-state UV−vis absorbance and (c) PL spectra of ZnS:Mn core NPLs (black) and ZnS:Mn/ZnS core/shell NPLs (red). PL decay of core NPLs (d) and core/shell NPLs (g) at various wavelengths. Three representative PL decay curves of core NPLs (e) and core/shell NPLs (h) for which detection wavelengths are set at 560, 600, and 640 nm, respectively. PL spectra of core NPLs (f) and core/shell NPLs (i) over decay time.

Taken together, we clearly see that the presence of the surface $Mn^{2+}$ ions results in a red-shift and broadening of the PL spectra of ZnS:Mn NPLs and a shortening of the decay lifetime. To disclose the relationship between the three $Mn^{2+}$ species (isolated, weakly coupled, and strongly coupled $Mn^{2+}$ ions) and their decay kinetics without the influence of the surface $Mn^{2+}$ ions, the ZnS:Mn/ZnS core/shell NPLs with low (0.1%, ≈ 4.9 $Mn^{2+}$ ions per NPL), intermediate (0.7%, ≈ 33 $Mn^{2+}$ ions per NPL), and high (4.9%, ≈ 233 $Mn^{2+}$ ions per NPL) doping levels were further prepared and considered. For all three differently doped samples, their PL decay lifetimes substantially increase after shell formation due to passivation of the surface $Mn^{2+}$ ions (Figure 6a−c). For comparison to the core-only samples we again used the kHz laser system. Here, these PL decay curves were fitted again with the triexponential function (see Eq. 1) and the time constants of the three $Mn^{2+}$ species are summarized in Table 2.

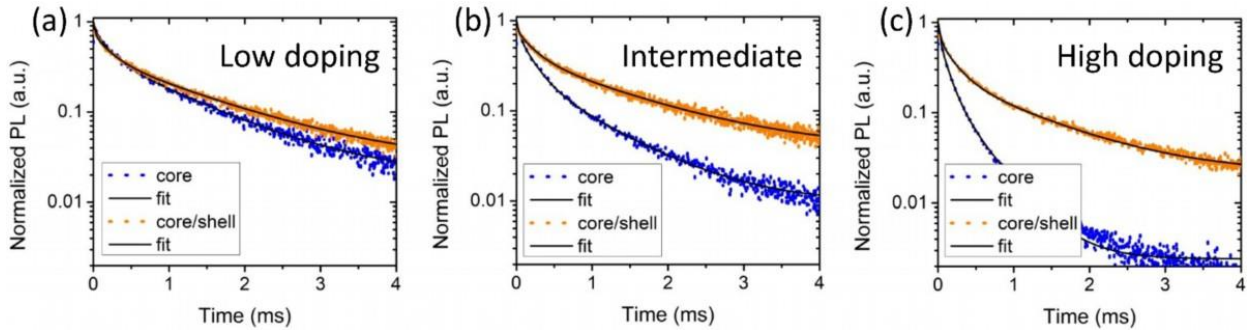

**Figure 6.** (a)−(c) PL decay curves of ZnS:Mn core NPLs (blue) and ZnS:Mn/ZnS core/shell NPLs (orange) with different doping levels.

For the low doping sample, where we expect predominant emission from isolated $Mn^{2+}$ ions, the measured long- and short-living component lifetimes increase to 2.46 ms and 315 µs, respectively, after shell formation, which is closer to the corresponding simulated lifetimes (3.2 ms and 370 µs).

Note that the measured 2.46 ms are already in very good agreement with the theory since the observation time of roughly 4 ms between two pulses is considerably short for such a long lifetime. Indeed, in the experiments with an optical chopper where longer observation times are possible (see experimental section) we found even longer lifetimes. Similar results are found for the intermediate doping sample where the emission from weakly coupled $Mn^{2+}$ ions should be predominant. When the surface $Mn^{2+}$ ions are passivated, their measured long- and short-living component lifetimes increase from 0.4−0.85 ms to 1.08−1.30 ms and from 114−196 μs to 206−226 μs, reaching the time scale of the simulated lifetimes. These findings reveal that the decay channels of the isolated and weakly coupled $Mn^{2+}$ ions are highly influenced by the surface $Mn^{2+}$ ions. However, the short-living component lifetime of the strongly coupled $Mn^{2+}$ ions keeps almost constant after shell formation, reflecting the independence of this decay channel on the surface effects.

**Table 2.** Time constants of the three $Mn^{2+}$ species measured for the core and core/shell NPLs and simulated by DFT calculations, respectively.

| $Mn^{2+}$ species | samples | long (ms) | short (μs) |
|---|---|---|---|
| isolated | core | 1.07 | 199 |
| | core/shell | 2.46 | 315 |
| | simulation | 3.20 | 370 |
| weakly coupled | core | 0.40–0.85 | 114–196 |
| | core/shell | 1.08–1.30 | 208–226 |
| | simulation | 1.55–3.16 | 220–330 |
| strongly coupled | core | - | 19–30 |
| | core/shell | - | 23–31 |
| | simulation | - | 50 |

CONCLUSIONS

In summary, we investigated the doping-level-dependent kinetics of the $Mn^{2+}$ $d-d$-related emission in ultrathin and atomically flat ZnS:Mn NPLs with the help of TRPL spectroscopy and DFT simulations. First, a qualitative picture of the interplay between slow, intermediate, and fast decay channels in the PL decays depending on the $Mn^{2+}$ doping level was established. A clear trend between the doping level increase (reduction of the average $Mn^{2+}-Mn^{2+}$ distance) and a reduction of the average PL lifetime was observed. Further, applying the core/shell approach (c-ALD) we found that surface $Mn^{2+}$ ions produce short-living red-shifted emissive states in our system that contribute to the broad PL spectrum and dominate in the long-wavelength shoulder. We fabricated a series of samples with suppressed surface ions contribution that opened the possibility to study the role of $Mn^{2+}-Mn^{2+}$ magnetic coupling. Reduced PL lifetimes with growing doping levels caused by the coupling were attributed to the progressive hybridization of $d$-orbitals of $Mn^{2+}$ ions in ZnS and lifting of the spin forbidden selection rules. Based on DFT simulations, we show that two emissive states with ~1–3 ms (long component) and ~50–300 μs (short component) lifetimes are present in the emission of $Mn^{2+}$ ions in ZnS NPLs due to different hybridization of Mn $d$-orbitals with the orbitals of the host material. Both components are inversely dependent on the $Mn^{2+}-Mn^{2+}$ distance. The origin of the slow decay channels is assigned to the long-living transitions ($p-d$-character transitions) of $Mn^{2+}$ ions whereby the intermediate decay channels are attributed to the short-living $s-p$-character transitions in $Mn^{2+}$ ions both in isolated and coupled ions. The component of ~50 μs is only associated with the strongly coupled $Mn^{2+}$ ions ($s-p$ transition) at the closest distance of 3.8 Å, implying aggregation or clustering of dopants. This explains the appearance of the three components in experimental samples in line with the scenarios of isolated, weakly- and strongly-coupled $Mn^{2+}$ ions. Proposed model allows potentially to form localized states with controlled lifetimes and emission wavelengths based on location-dependent and $Mn^{2+}-Mn^{2+}$ distance-dependent effects on PL in ZnS NPLs and serves as guiding tool for the assessment of the doping level, homogeneity and surface contribution of $Mn^{2+}$-doped and in general doped nanosystems.

EXPERIMENTAL SECTION

**Chemicals.** Zinc chloride (97+%), octylamine (99+%), and methanol (99.8%) were purchased from Acros. Sulfur powder (99.998%), manganese (II) acetate (98%), oleylamine (70%), ammonium sulfide (($NH_4)_2S$, 20 wt% in $H_2O$), formamide (99+%), and N-methylformamide (NMF, 99%) were ordered from Sigma-Aldrich. Toluene (99.5%), isopropanol (99.7%), and hexane (95%) were purchased from VWR. Acetone (99%) was purchased from Th. Geyer Chemicals. All chemicals were used without further purification.

**Synthesis of ZnS:Mn NPLs.** ZnS:Mn NPLs were synthesized according to a reported recipe.[30] Briefly, 0.15 mmol (20.4 mg) of $ZnCl_2$, 0.45 mmol (14.4 mg) of sulfur powder, and a certain amount of $Mn(OAc)_2$ were dissolved in a mixture of 5 mL of octylamine and 10 mL of oleylamine in a 25 mL three-necked flask. The mixture was purged with nitrogen at 100 °C for 30 min under vigorous stirring. The reaction solution was then heated to 150 °C for 6 hours with magnetic stirring under nitrogen flow. After the reaction, the mixed solution was naturally cooled to room temperature. To remove excess ligands and unreacted precursors, the reaction solution was mixed with acetone/isopropanol (v:v = 1:1). Then the mixture was shaken well and centrifuged for 10 min at 9000 rpm. Finally, the supernatant was discarded, and the precipitation was re-dispersed into 3 mL of hexane or toluene for further characterization. For the synthesis of ZnS:Mn NPLs with different $Mn^{2+}$ concentrations, the amount of $Mn(OAc)_2$ was varied while other parameters remain unchanged.

**Synthesis of ZnS:Mn/ZnS Core/Shell NPLs.** ZnS:Mn/ZnS core/shell NPLs were prepared by using c-ALD method.[36] Briefly, 200 μL of as-synthesized ZnS:Mn NPLs was purified by centrifugation, and the precipitation was then re-dispersed into 1 mL of hexane in a 5 mL glass vial. 1 mL of NMF and 22 μL of aqueous $(NH_4)_2S$ solution (20 wt% in $H_2O$) were added into the vial. The two-phase mixture was stirred vigorously until complete phase transfer of NPLs from hexane to NMF. The upper phase (hexane) was discarded. The lower phase (NMF) containing $S^{2-}$-coated NPLs was rinsed two times with hexane, followed by centrifugation and re-dispersion in 1 mL of NMF. To deposit a layer of $Zn^{2+}$ ion, 120 μL of 0.1 M $ZnCl_2$ solution in formamide was added, and the mixture was stirred vigorously for 30 s. Finally, the NPLs were purified by centrifugation and re-dispersed in 1 mL of NMF. The above procedure was repeated three times.

**Steady-state UV−vis Absorbance, PL, and TRPL Spectroscopy.** The steady-state UV−vis absorbance and PL spectra were obtained with a PerkinElmer Lambda 25 two-beam spectrometer and a Horiba Fluoromax-4 spectrometer, respectively. The TRPL measurements were performed with a Picoquant FluoTime 300 fluorescence spectrometer. Most of the TRPL experiments were performed with a low repetition rate supercontinuum laser (NKT Photonics, SuperK COMPACT) connected to a second harmonic generation unit (NKT Photonics, SuperK EXTEND deep UV) providing 285 nm pulses of 1−2 ns pulse length and a repetition rate of 240 Hz. To obtain a sufficient signal-to-noise ratio the bandwidth of the spectrometer was set to 27 nm centered at the maximum of the $Mn^{2+}$ emission band. Using a PMA Hybrid 07 photomultiplier by Picoquant as detector the integration time was set to 1 hour. The measurements concerning the surface effect on the $Mn^{2+}$ emission (summarized in Figure 5) needed a different excitation scheme. Here, a high repetition rate supercontinuum laser (NKT Photonics, SuperK FIANIUM FIU6 was connected to the second harmonic generation unit now providing 285 nm pulses of 10−20 ps pulse length at 78.5 MHz repetition rate. The time resolution was obtained by an optical chopper wheel transmitting and blocking the beam with 50 Hz repetition rate at 50% duty cycle. The chopper experiment provides a much higher time averaged laser power and thus a higher count rate. This allows for wavelength- and time-resolved fluorescence measurements (Figure 5d and g) with a much lower bandwidth (2.5 nm) and longer measurement time (10 ms = 50% of 20 ms modulation period). Those measurements enable to observe the transient fluorescence spectra in Figure 5(f, i) showing the wavelength shifts caused by the surface $Mn^{2+}$ emission. However, note that the chopper experiments have – on the one hand – two major drawbacks: (i) Fast components are strongly affected by the limited time resolution caused by the geometrical relation of beam diameter and aperture of the chopper wheel. (ii) The quasi CW excitation during the open cycle of the chopper predominantly populates long-living $Mn^{2+}$ states compared to short-living states. This is a direct consequence of the long excitation cycle of 10 ms during that long-living states reach equilibrium between excitation and recombination at a higher population than short-living states. For this reason we performed most of the experiments with the pulsed laser. On the other hand, the predominant population of the long-living states and the prolonged observation time makes the chopper experiments more sensitive to the long-living emission components which are very useful for addressing the single quasi isolated $Mn^{2+}$ ions. The samples for the optical measurements were prepared by adding a few microliters of the NPL solution into 3 mL of hexane in quartz vessels with an optical path length of 10 mm.

**High-resolution Transmission Electron Microscopy (HRTEM).** TEM and HRTEM images were obtained using a Philips CM300 UT microscope operated at an acceleration voltage of 200 kV. High angle annular dark-field STEM images were further acquired using a probe corrected Jeol JEM-ARM200F NeoARM electron microscope at 200 kV acceleration voltage. Samples for the HRTEM analysis were prepared by drop-casting 10 μL of the diluted NPL dispersion onto carbon-coated copper grids.

**Electron Paramagnetic Resonance (EPR) Spectroscopy.** EPR measurements were recorded at 100 K on a Bruker EMX CW-micro X-band EPR spectrometer equipped with an ER4119HS high-sensitivity resonator, with a microwave power of Ca 6.9 mW, modulation frequency of 100 kHz, and amplitude of 5 G. The EPR spectrometer was equipped with a temperature controller and liquid $N_2$ cryostat for low-temperature measurements. For each measurement, 2 mL of ZnS:Mn NPLs with different $Mn^{2+}$ doping levels (in hexane) was used. For calculation of $g$ values, the equation $h\nu = g\beta B_0$ was used with $\beta$, $B_0$, and $\nu$ being the Bohr magneton, resonance field, and frequency, respectively. Calibration of the $g$ values was performed by using a DPPH standard ($g = 2.0036 \pm 0.0004$). The EPR spectra of the samples were simulated by using the software package Easyspin implemented in MATLAB.

**DFT Calculations.** DFT calculations have been carried out within the local density approximation (LDA). The slab was constructed as 3×3×4 repetition of a ZnS bulk supercell with 8 atoms and thus containing 288 atoms. Structural relaxation with stress minimization has been performed using the default convergence values of the Quantum espresso package followed by ground state calculations to generate the Kohn-Sham energies and wave functions. Note that the relative position of the $Mn^{2+}$ $d$-level with respect to the ZnS bulk VBM and CBM are not accurate at the DFT-LDA level due to the known lack of correlation. This could have been corrected with post-DFT methods such as the GW approximation. However, due to the large simulation cell, this type of calculations could not be performed. It is also worth noting that the previous problem is not supposed to influence the lifetime calculation. The lifetime is mainly wave function dependent only through transition dipole moment (electron-hole $M_{if}$). In most cases, DFT and GW wave functions are nearly identical.

ASSOCIATED CONTENT

**Supporting Information**

TEM images of core and core/shell NPLs, steady-state UV−vis absorbance and PL spectra, Fitting parameters, EPR spectra, transient PL spectra of ZnS:Mn NPLs, XRD pattern of core/shell NPLs.

AUTHOR INFORMATION

**Corresponding Author**

* rostyslav.lesyuk@uni-rostock.de

**Author Contributions**

The manuscript was written through contributions of all authors. All authors have given approval to the final version of the manuscript.

**Notes**

The authors declare no competing financial interest.

ACKNOWLEDGMENT

L.D. thanks the China Scholarship Council (CSC) for financial support. We are thankful for the help of Andreas Kornowski (†), University of Hamburg. We thank Stefan Werner (University of Hamburg) for TEM measurements. A.M., T.K., G.B. and L.D. are grateful for funding by the Deutsche Forschungsgemeinschaft via the Research Training Group "Nanohybrid" (GRK2536/1). C.K. is grateful for the support by European Regional Development Fund of the European Union (GHS-20-0035/P000376218, GHS-20-0036/P000379642). R.L. and C.K. thank the German Research Foundation (Deutsche Forschungsgemeinschaft) for funding the supporting collaborative research center LiMatI (SFB 1477) and for funding the Jeol JEM-ARM200F NeoARM TEM (DFG INST 264/161-1 FUGG).